\documentclass{iaus}
\usepackage{graphicx}
\usepackage{subfigure}

\title[MASSCLEAN -- Description, Tests, and Results] 
{\LARGE MASSCLEAN \\ \bitlarger -- MASSive CLuster Evolution and ANalysis Package -- \\ \Large Description, Tests, and Results}

\author[Popescu \& Hanson]   
{Bogdan Popescu
 \and M. M. Hanson}

\affiliation{Department of Physics, University of Cincinnati, Cincinnati, OH, USA \\ email: {\tt popescb@mail.uc.edu, margaret.hanson@uc.edu}}

\pubyear{2009}
\volume{266}  
\pagerange{1--4}
\jname{Star Clusters}
\editors{Richard de Grijs \& Jacques R. D. L\'epine, eds.}
\begin{document}

\maketitle

\begin{abstract}
MASSCLEAN is a new, sophisticated and robust stellar cluster image and photometry simulation package. This
package is able to create color-magnitude diagrams and standard FITS images in any of the traditional optical and
near-infrared bands based on cluster characteristics input by the user, including but not limited to distance, age, mass,
radius and extinction. At the limit of very distant, unresolved clusters, we have checked the integrated colors created in
MASSCLEAN against those from other simple stellar population (SSP) models with consistent results. Because the algorithm
populates the cluster with a discrete number of tenable stars, it can be used as part of a Monte Carlo Method to derive
the probabilistic range of characteristics (integrated colors, for example) consistent with a given cluster mass and age.
We present the first ever mass dependent integrated colors as a function of age, derived from over 100,000 Monte
Carlo runs, which can be used to improve the current age determination methods for stellar clusters.
\end{abstract}
\firstsection 
\vspace*{-0.4 cm}              
\section{Introduction}

{\it Star clusters are the building blocks of galaxies}. In order to analyze the known population of star clusters, to search for new clusters, and to derive the selection effects of current NIR survey searches for Milky
Way clusters we created a simulation package.

The \texttt{MASSCLEAN} package\footnote{The package is freely available under GNU General Public License at: 
\texttt{http://www.physics.uc.edu/\textasciitilde bogdan/massclean/}
}  (\cite[Popescu \& Hanson 2009]{masscleanpaper}; \cite[Hanson \& Popescu 2007]{aas211}, \cite[2008]{iaus250}, \cite[2009]{aas214})
is built using numerous well established theoretical and empirical models for stars and stellar clusters, starting with the Kroupa-Salpeter IMF for stellar mass distribution (\cite[Kroupa 2002]{Kroupa2002}; \cite[Salpeter 1955]{Salpeter1955}). 
The user has the option to choose between two stellar evolutionary models: the Geneva Models (\cite[Lejeune \& Schaerer 2001]{geneva1}) or the Padova Models (\cite[Marigo et al. 2008]{padova2008}).  The extinction model is based on \cite[Cardelli, Clayton and Mathis (1989)]{ccm89}, and the King Model is used for the spatial distribution (\cite[King 1962]{King1962}). An optional stellar field can be generated using the "SKY" Model (\cite[Bahcal \& Soneira 1984]{bahcall1984}). \texttt{MASSCLEAN} is built in a modular way and is able to combine observational data and simulations using an unprecedented level of accuracy.

\vspace*{-0.5 cm}
\section{Images -- Galactic and Extragalactic Simulations}

\begin{figure}[htb]
\vspace*{-0.1 cm}
\begin{center}
\subfigure[Real Image (2MASS)] {\includegraphics[width=0.33\textwidth]{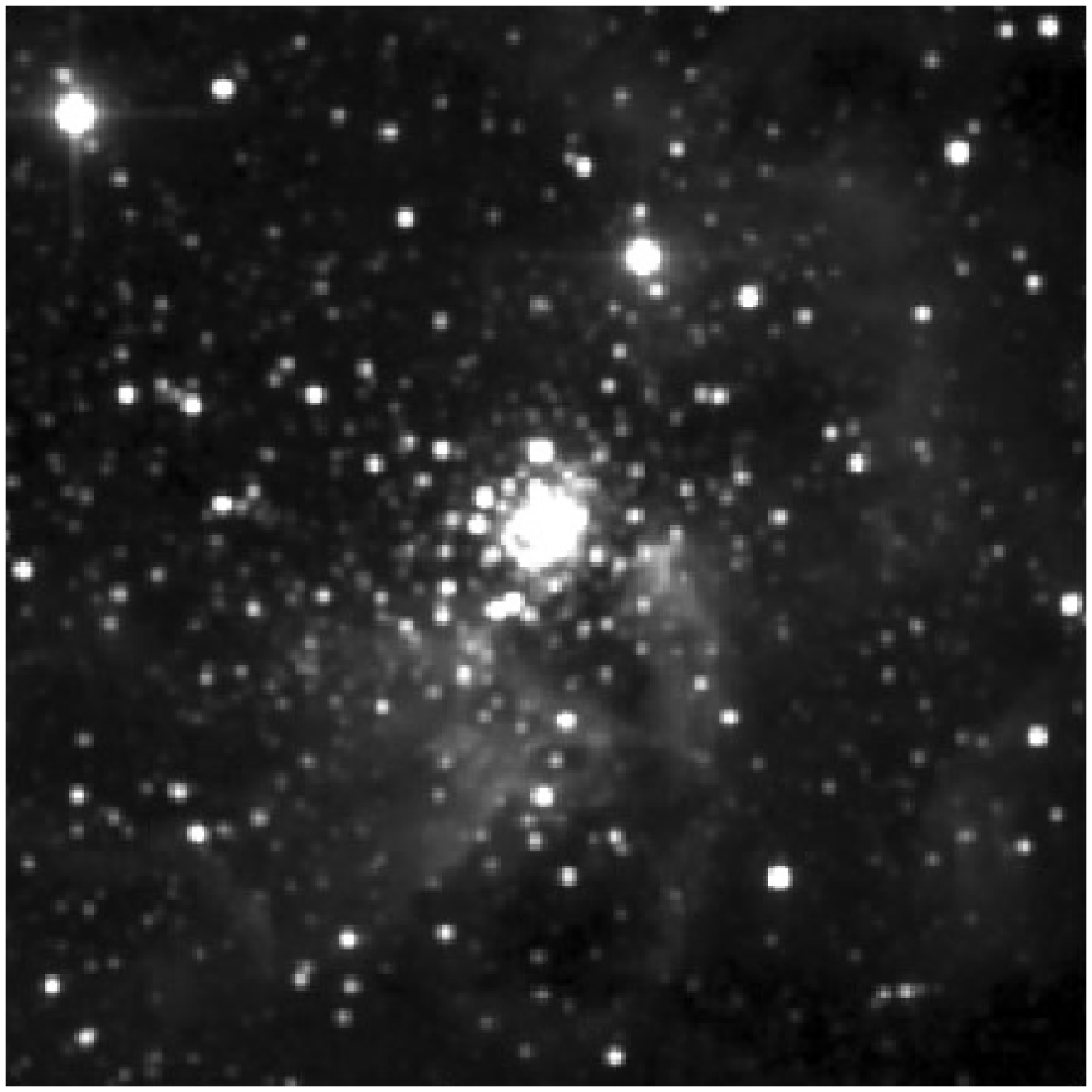} } 
\subfigure[Simulated Image] {\includegraphics[width=0.33\textwidth]{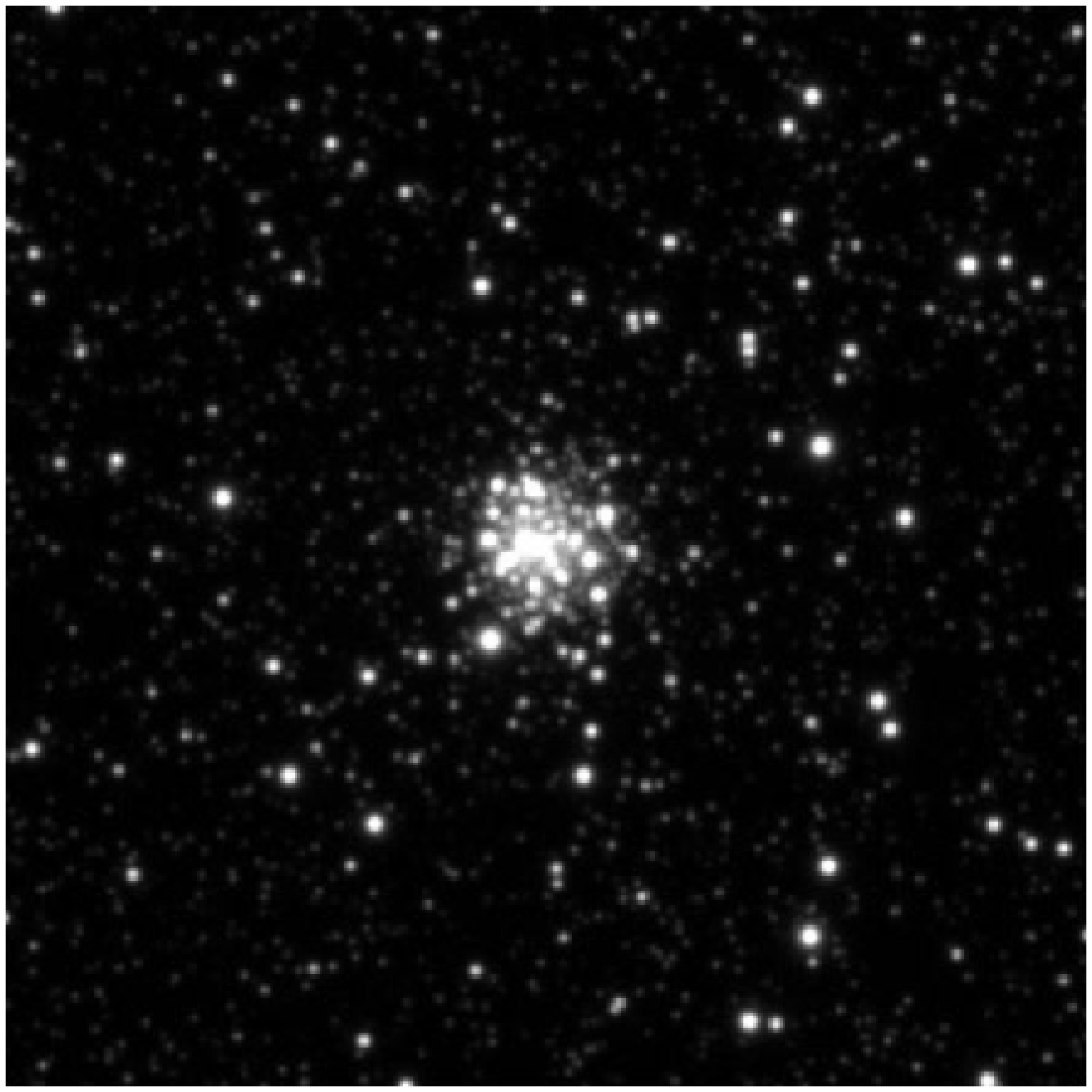} }
\vspace*{-0.25 cm}
 \caption{{\textit NGC 3603} J-Band Images ($4.4 \times 4.4$ arcmin).}
   \label{fig1}
\end{center}
\end{figure}

\texttt{MASSCLEAN} generates FITS images, using {\sc SkyMaker} (\cite[Bertin 2009]{ber09}).
The simulated $J$-band image for {\textit NGC 3603} is presented in Fig.\,\ref{fig1} (b), and the 2MASS\footnote{\scriptsize This publication makes use of data products from the Two Micron All Sky Survey, which is a joint project of the University of Massachusetts and the Infrared Processing and Analysis Center/California Institute of Technology, funded by the National Aeronautics and Space Administration and the National Science Foundation.} image is presented in Fig.\,\ref{fig1} (a).


\vspace*{-0.5 cm}   

\begin{figure}[htb]
\vspace*{-0.1 cm}
\begin{center}
\subfigure[$log(age/yr)=6.0$] {\includegraphics[width=0.29\textwidth]{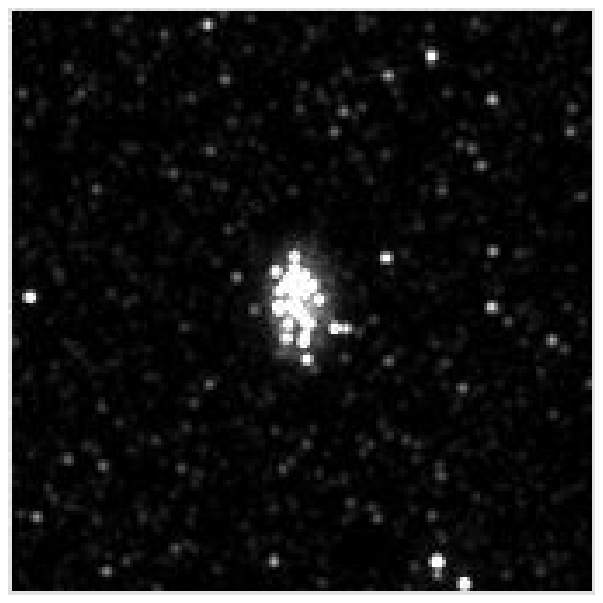} }
\subfigure[$log(age/yr)=8.0$] {\includegraphics[width=0.29\textwidth]{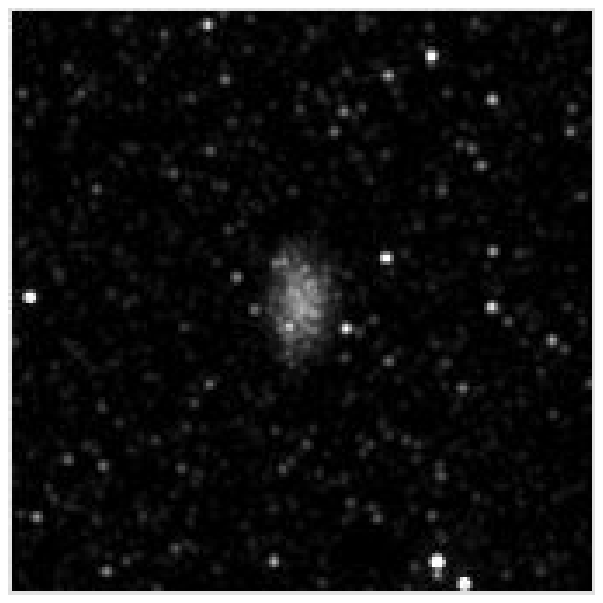} }
\vspace*{-0.25 cm}
 \caption{$10^5$ $M_{\odot}$ simulated cluster at the distance of $M31$ and the
resolution of {\it HST ACS}.}
   \label{fig2}
\end{center}
\end{figure}

In the Fig.\,\ref{fig2} we present a simulated $10^5$ $M_{\odot}$ cluster at the distance of $M31$ and the resolution of {\it HST ACS}. The images show the clusters at ages of 1  and 100 million years.

\vspace*{-0.25 cm}   

\begin{figure}[htb]
\vspace*{-0.1 cm}
\begin{center}
\subfigure[$V$-Band Image] {\includegraphics[width=0.28\textwidth]{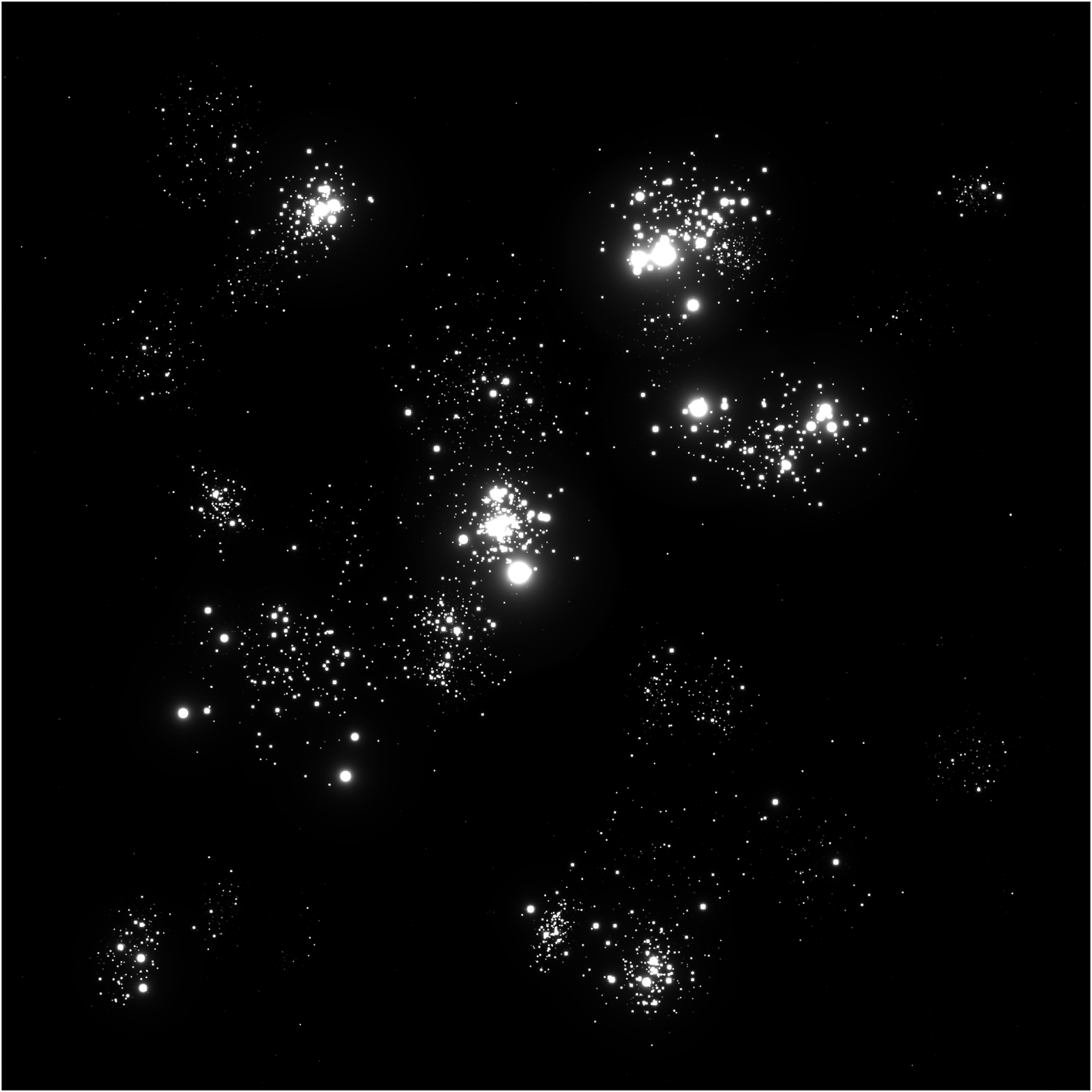} } 
\subfigure[$K$-Band Image] {\includegraphics[width=0.28\textwidth]{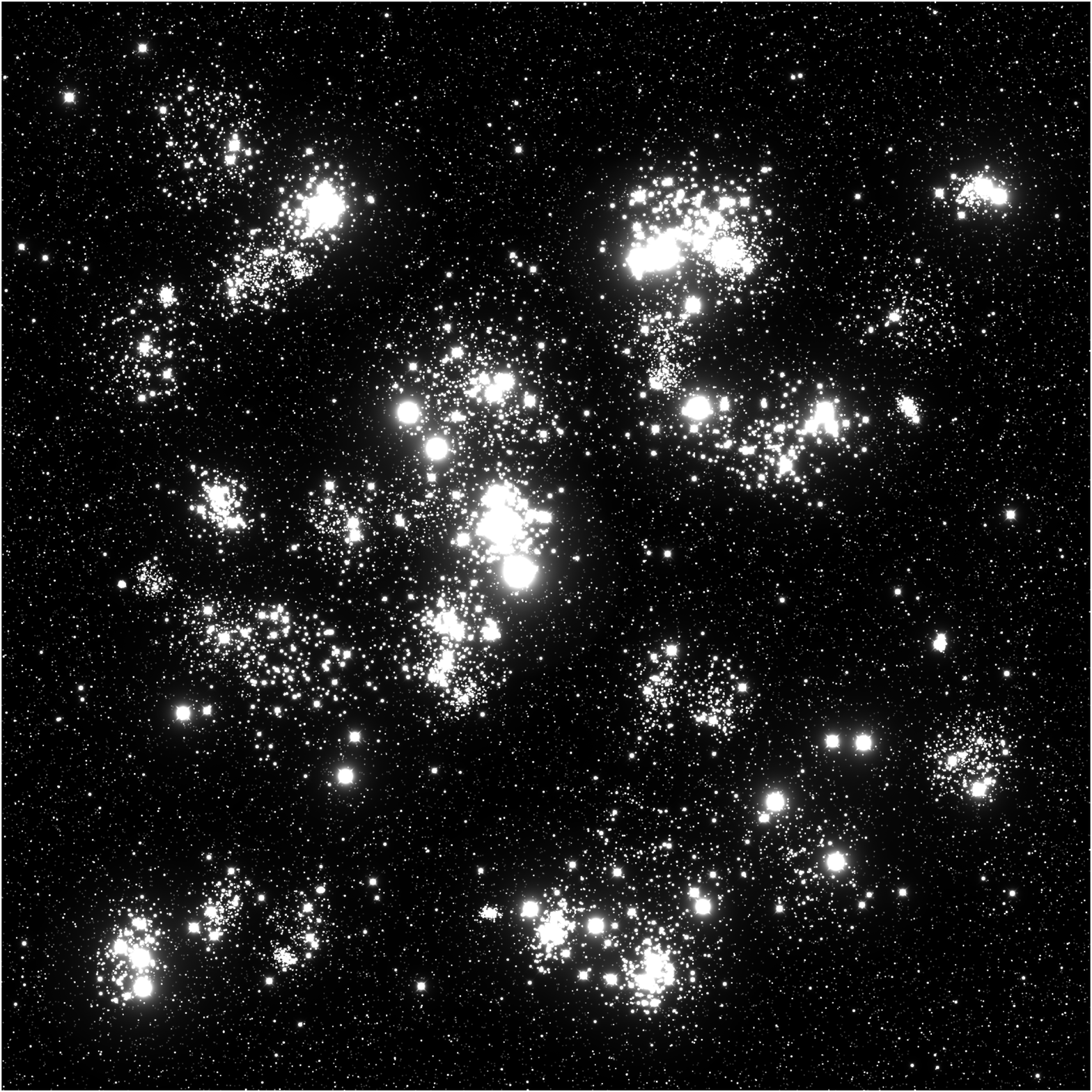} } 
\subfigure[Positions] {\includegraphics[width=0.28\textwidth, bb=83 179 542 638]{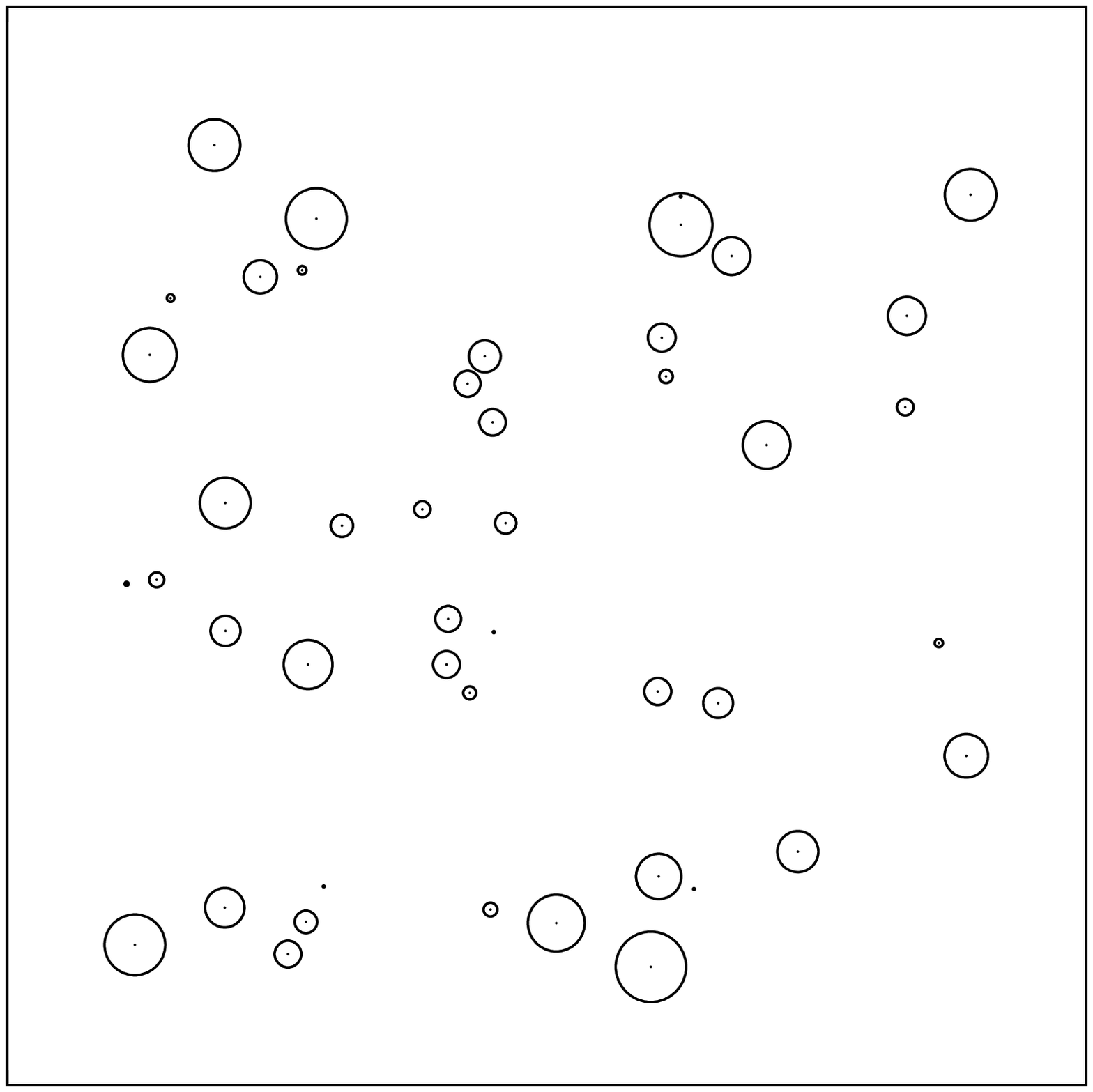} }
\vspace*{-0.25 cm}
 \caption{45 simulated clusters}
   \label{fig3}
\end{center}
\end{figure}

We are using \texttt{MASSCLEAN} to help us derive the selection effects of current NIR survey searches for Milky
Way clusters (\cite[Hanson et al. 2009]{iaurio}; Hanson et al., in prep.). As a proof of concept, we present an example image simulation of 45 randomly generated clusters of various age, mass, concentration, and extinction. The $V$-band image is shown in Fig.\,\ref{fig3} (a), and the $K$-band image is shown in Fig.\,\ref{fig3} (b) (note, no stellar field is included). The positions of the clusters along with the scaled angular size of the clusters' core is shown in Fig.\,\ref{fig3} (c).

\vspace*{-0.75 cm}
\section{Integrated Colors and Ages of Clusters}

\begin{figure}[htb]
\vspace*{-0.1 cm}
\begin{center}
\subfigure[Comparison with modern SSP codes] {\includegraphics[width=0.44\textwidth]{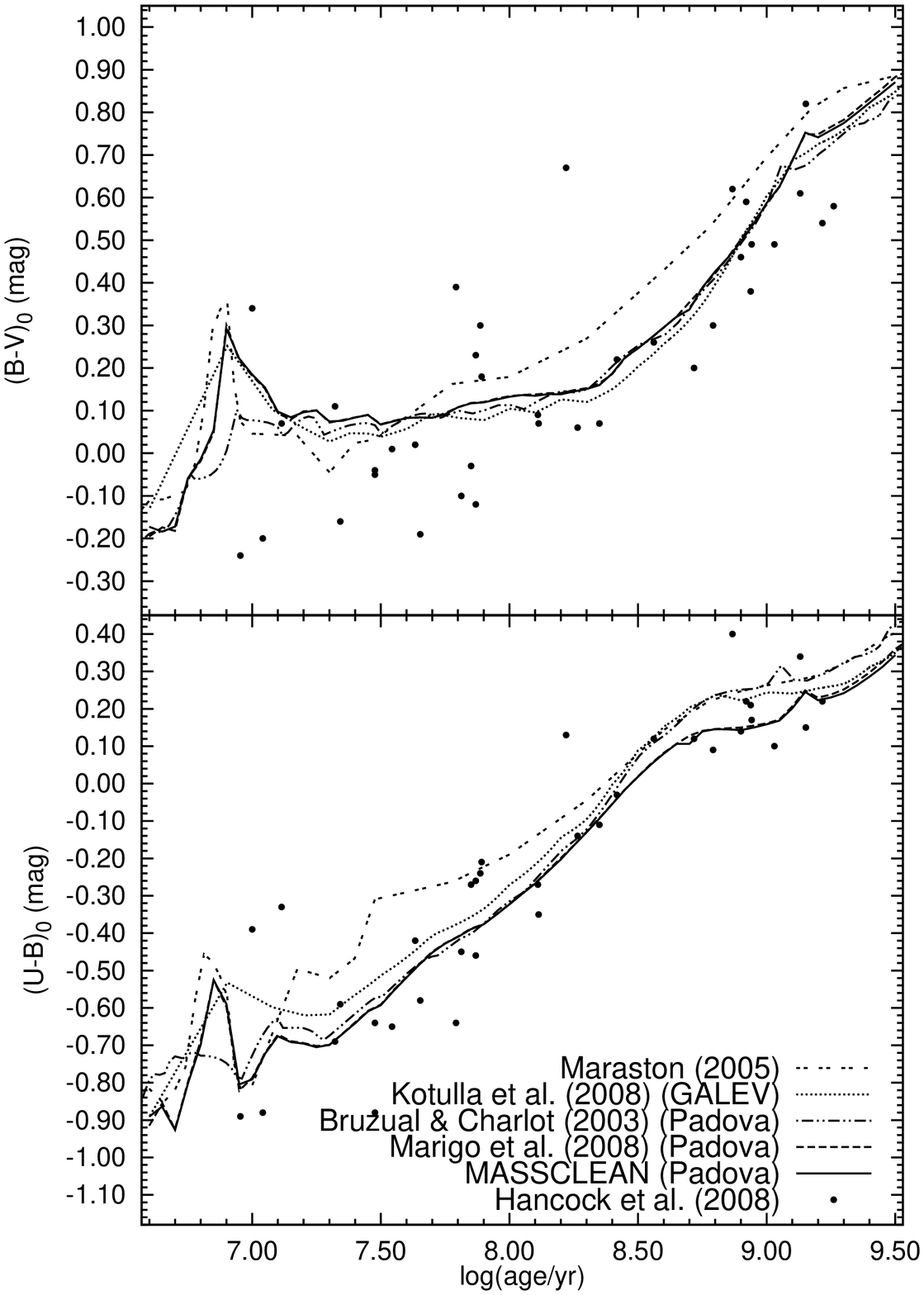} }
\subfigure[Mass dependent integrated colors] {\includegraphics[width=0.44\textwidth]{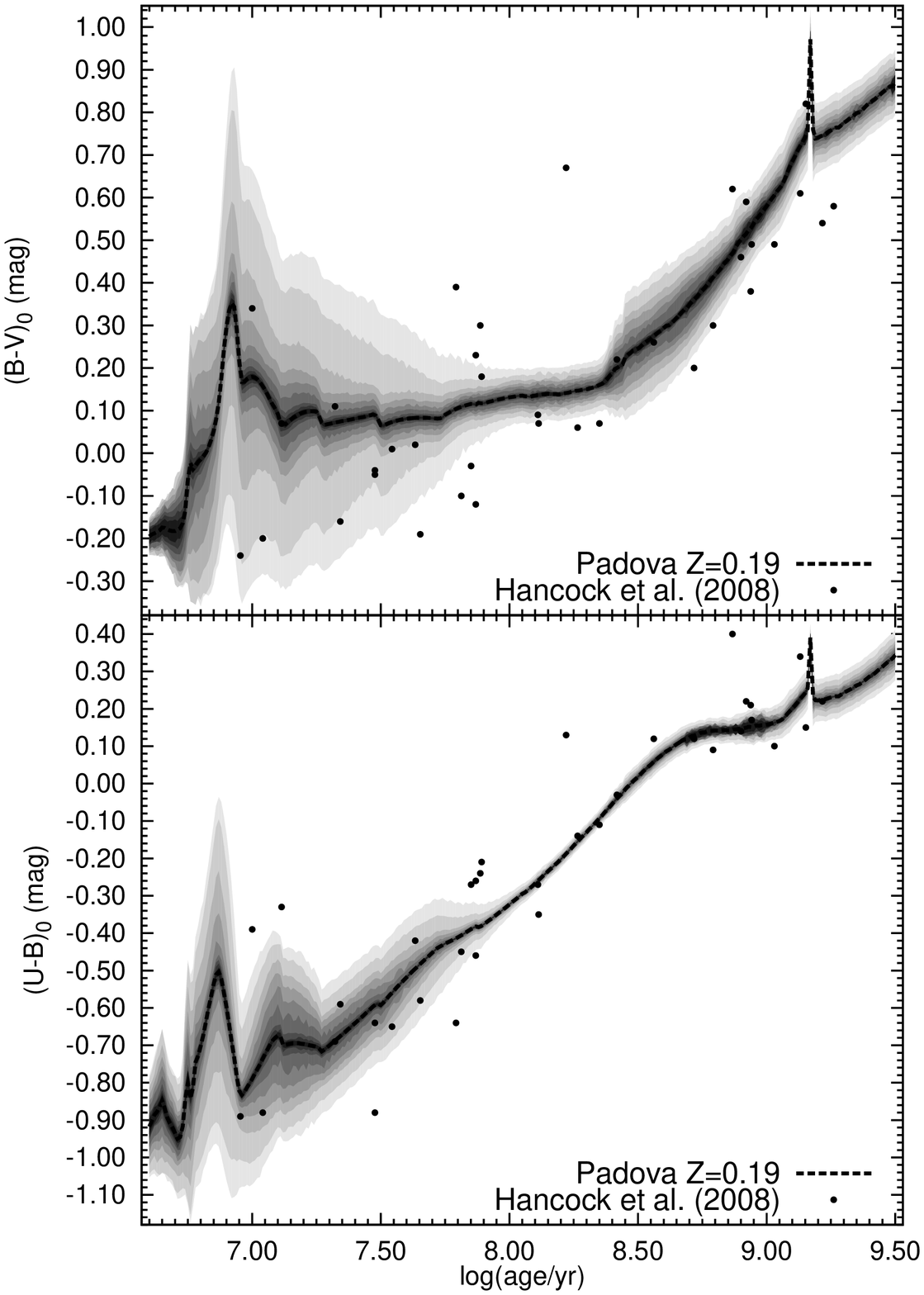} }
\vspace*{-0.25 cm}
 \caption{\texttt{MASSCLEAN} integrated colors}
   \label{fig4}
\end{center}
\end{figure}

MASSCLEAN follows the evolution of every star in the cluster, providing full photometry at all ages. We have
tested that at the infinite mass limit the integrated colors are consistent with standard SSP models.
A $log(age/yr)$ step of $0.05$ has been used to allow comparison with all other models.
The integrated colors generated by \texttt{MASSCLEAN} using Padova 2008 evolutionary models for solar metalicity are presented in Fig.\,\ref{fig4} (a). The results from other SSP codes (\cite[Marigo et al. 2008]{padova2008}; \cite[Bruzual \& Charlot 2003]{bruzual2003}; \cite[Maraston 2005]{maraston2005}; 
\cite[Kotulla et al. 2009]{galev2009}) are also displayed, along with the Milky Way clusters from \cite[Hancock et al. 2008]{hancock}.  The \texttt{MASSCLEAN} integrated colors follow the \cite[Marigo et al. 2008]{padova2008} since they both use the same evolutionary models for their input.

\vspace*{-0.25 cm}   

Of course, real star clusters have a finite mass.  They do not have fully populated isochrones at all ages and their integrated colors may not always correspond to the ones computed in the infinite mass limit (\cite[Bruzual 2002]{bruzual2002}; \cite[Cervi\~no \& Luridiana 2004]{cervino2004}; \cite[Cervi\~no \& Luridiana 2006]{cervino2006}; \cite[Fagiolini et al. 2007]{fagiolini2007}; \cite[Lan{\c c}on \& Mouhcine 2000]{lancon2000}; \cite[Lan{\c c}on \& Fouesneau 2009]{lancon2009}).
 The dispersion of integrated
colors  ($1 \sigma$) due to different values of mass and fluctuations in the IMF is presented in Fig.\,\ref{fig4} (b). 
$(B-V)_{0}$ vs. $log(age/yr)$ and $(U-B)_{0}$ vs. $log(age/yr)$ are presented for a few values of cluster's mass. 
The step in $log(age/yr)$ is $0.01$.
The color dispersion for clusters with mass in the $2 \times 10^{3}$ $M_{\odot}$
-- $2.5 \times 10^{5}$ $M_{\odot}$ range is based on about {\it 100,000 Monte Carlo runs} (which corresponds to about $15$ \textit{million
cluster models}).  We present \textit{for the first time} {\it mass dependent integrated
colors for clusters}. The entire set of mass dependent integrated colors using both Padova and Geneva models, and different metalicities will be publicly available soon (Popescu \& Hanson, in prep.).

\begin{figure}[htb]
\vspace*{-0.1 cm}
\begin{center}
\subfigure[U-B colors as a function of age.] {\includegraphics[angle=270,width=0.44\textwidth]{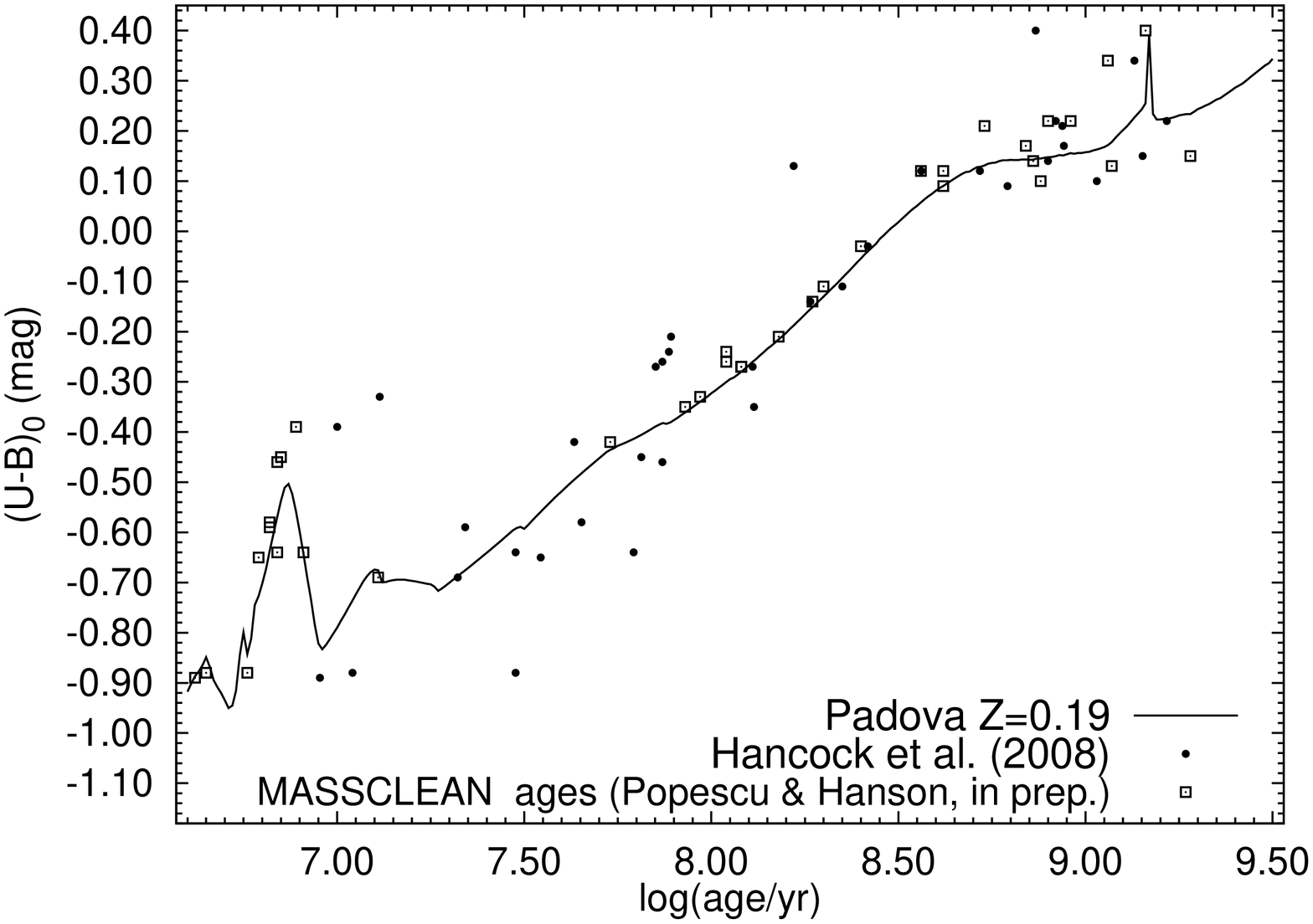} }
\subfigure[B-V colors as a function of age.] {\includegraphics[angle=270,width=0.44\textwidth]{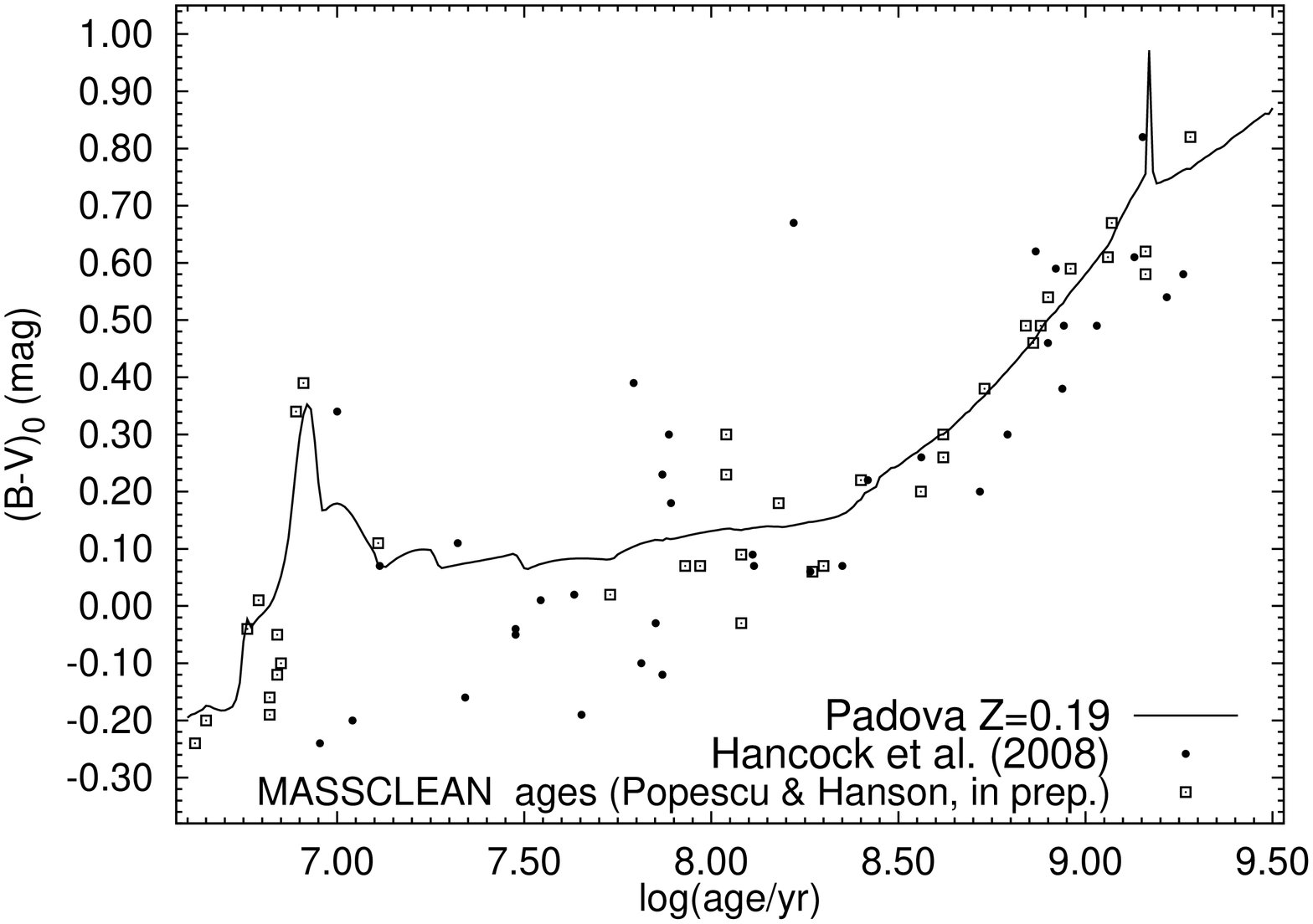} }
\vspace*{-0.25 cm}
 \caption{Real clusters, with masses of just a few thousand solar mass, are expected to be redder than SSP models predict using an 'infinite mass assumption' particularly in the age range 10$^6 - 10^8$. We are testing this result using integrated colors of Milky Way clusters with well constrained ages (Popescu \& Hanson 2009, in prep.).}
   \label{fig5}
\end{center}
\end{figure}

Age determination for clusters using integrated colors is usually based only on the infinite mass limit.
Using our mass dependent integrated colors and magnitudes, we expect to provide a more realistic model.  As an initial demonstration, we present new age determination based on \texttt{MASSCLEAN} integrated colors for numerous
Milky Way clusters in Fig.\,\ref{fig5}. (Popescu \& Hanson, in prep).

\vspace*{-0.75 cm}
\section{Summary}

We have developed a sophisticated cluster image and photometry 
simulation package which we are using to test against current search algorithms. We predict that
clusters over certain age ranges (particularly when the red supergiant phase is just starting
out) and clusters with low stellar densities (even though possibly of very high total
mass) are being missed by the current searches, particularly when combined with high
extinction.
Our mass dependent integrated colors for clusters should provide a better description for 
known clusters and a more accurate age determination.

\end{document}